\documentclass{scrartcl}

\usepackage{amsmath}
\usepackage{physics}
\usepackage{amssymb}
\usepackage{amsthm}
\usepackage{xcolor}
\usepackage{graphicx}
\usepackage{comment}
\usepackage{tikz-cd}
\usepackage{pgfplots}
\usepackage{varwidth}
\usepackage{hyperref}
\usetikzlibrary{shapes,arrows,patterns}

\tikzstyle{block} = [rectangle, draw, fill=MPIblue!20, text width=2em, text centered, rounded corners, minimum height=2em]

\definecolor{MPIdarkblue}{cmyk}{1,0.5,0.0,0.72}
\definecolor{MPIblue}{cmyk}{0.22,0.11,0.0,0.36}
\definecolor{MPIbeige}{cmyk}{0,0,0.08,0.23}
\definecolor{MPIred}{cmyk}{0.29,0.93,0.96,0.05}
\definecolor{MPIgrey50}{cmyk}{0.07,0.03,0.07,0.0}
\definecolor{MPIgreen}{RGB}{0,119,117}
\definecolor{MPIgreen50}{RGB}{128,187,186}
\definecolor{MPIgrey}{RGB}{221,222,214}
\definecolor{MPIred}{RGB}{217,63,39}
\definecolor{MPIgreygreen50}{RGB}{174,205,200}
\definecolor{MPIgreengrey50}{RGB}{119,179,176}
\definecolor{MPIgrey50green50}{RGB}{183,213,210}

\definecolor{MPIblack}{RGB}{0,0,0}
\definecolor{MPIwhite}{RGB}{255,255,255}

\definecolor{MPIdarkgrey}{RGB}{100,89,89}
\definecolor{MPIdarkyellow}{RGB}{239,208,85}

\title{Quantum Measurement and Objective Classical Reality}

\author{Vishal Johnson$^{1,2,*}$, Philipp Frank $^{1}$, Torsten En{\ss}lin$^{1,2}$ \\
{\small$^{1}$ Max Planck Institute for Astrophysics, Garching}\\
{\small$^{2}$ Ludwig-Maximilians-Universit{\"a}t, Munich}\\
{\small$^{*}$ Correspondence: vishal@mpa-garching.mpg.de}}

\date{\today}

\begin{document}
	\maketitle
	\begin{abstract}
		We explore quantum measurement in the context of Everettian unitary quantum mechanics and construct an explicit unitary measurement procedure. We propose the existence of prior correlated states that enable this procedure to work and therefore argue that correlation is a resource that is consumed when measurements take place. It is also argued that a network of such measurements establishes a stable objective classical reality.
	\end{abstract}
	\begin{center}
	\end{center}

	\section{\label{sec:introduction} Introduction}
	The usual treatment of quantum mechanics is to consider two distinct kinds of dynamics for the quantum wavefunction, reversible unitary dynamics of the Schr{\"o}dinger equation and irreversible wavefunction collapse. Similar to the approach of Hugh Everett \cite{everett} we assume a completely unitary dynamics for quantum mechanics and explore its consequences. The aim is to explore the extent to which the results of quantum experiments, and thereafter our own subjective experience of our world, can be explained starting from a hypothesis of unitarity (or equivalently, reversible dynamics).
	
	In a related publication \cite{measurement-unitary} we explain the details of how to define a measurement procedure within the context of unitary quantum mechanics. The current article explores a particular instructive example in detail and sets it in the context of explaining the objective classical reality we are familiar with.
	
	A consequence of the assumption of unitarity is that redundant information about the measurement is required in order to correct for the inevitable environmental influence. As a result we propose the existence of a "correlated environment" that provides this information. This is related to "spectrum broadcast structures" described in \cite{spectrum-broadcast}. We thereafter explain the emergence of an objective classical reality following the arguments in \cite{zurek-classical}.
	
	The article is structured as follows: The limits that unitarity of quantum mechanics places on the measurement procedure and its consequences are explored in section \ref{sec:unitarity}. The motivation of a "correlated environment" and explanation of how it is a conserved resource follows in section \ref{sec:resource}. We then explain how these considerations help explain the familiar objective classical reality in section \ref{sec:objective}. A discussion follows in section \ref{sec:discussion}.
	
	\section{\label{sec:unitarity} Limits of Unitary Quantum Mechanics}
	In this section we discuss how the assumption of unitary quantum mechanics limits the dynamics of quantum measurement. The arguments presented here are quite similar to those discussed in \cite{everett}.
	
	Consider a quantum system being measured,
	\begin{equation} \label{eq:signal-spin}
		\ket{\psi}_\mathrm{s}=\psi_\uparrow\ket{\uparrow}_\mathrm{s} + \psi_\downarrow\ket{\downarrow}_\mathrm{s},
	\end{equation}
	where we use the mental picture of an electron spin being measured, which is labelled s for "signal". The electron spin could be either in the $\ket{\uparrow}_\mathrm{s}$ or $\ket{\downarrow}_\mathrm{s}$ state and generically it is in a superposition such as in equation \eqref{eq:signal-spin}.
	
	The quantum system is measured by an observer. As we assume that quantum mechanics is a complete description of our universe, the observer is also described using a quantum state. Let us assume that the observer is in the arbitrary state
	\begin{equation} \label{eq:observer-initial}
		\ket{\phi}_\mathrm{o}=\phi_\uparrow\ket{\uparrow}_\mathrm{o} + \phi_\downarrow\ket{\downarrow}_\mathrm{o}.
	\end{equation}
	The states $\ket{\uparrow, \downarrow}_\mathrm{o}$ for the observer indicate its physical configurations that correspond to recording\footnote{Due to time reversibility, it is not required to label the states with a time label; the state of the observer when the signal is measured corresponds to a well defined state when the measurement procedure is begun.} a measurement of $\ket{\uparrow, \downarrow}_\mathrm{s}$, respectively, for the signal. A specific initial state is assumed in usual approaches to Everettian quantum mechanics and this leads, implicitly or explicitly, to the issue of preferred bases \cite{Hemmo2022-HEMTPB}. In our approach, an arbitrary state for the observer is assumed (equation \eqref{eq:observer-initial}) so that the observer need not be in any predetermined initial state to allow the measurement procedure to take place.  Nevertheless, our approach does not completely resolve the issue as the basis dependence comes in implicitly at the level of the Hamiltonian (or unitary) enabling the measurement procedure; this is discussed in section \ref{sec:discussion}.
	
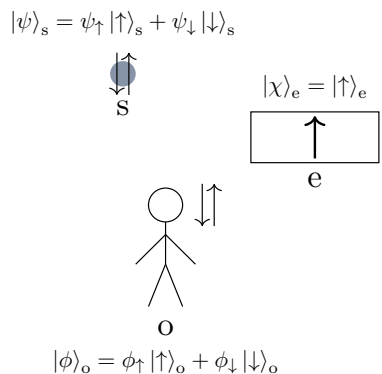
\begin{figure}[h]
		\centering
		\resizebox{5cm}{!}{\begin{tikzpicture}[auto]			
				\fill[MPIblue] (0,0) circle (0.15);
				\node at (0,-0.4) {s};
				
				\draw (0.5,-1.55) circle (0.2);
				\draw (0.5,-1.75) -- +(0,-0.5);
				\draw (0.5,-1.9) -- +(0.35,-0.35);
				\draw (0.5,-1.9) -- +(-0.35,-0.35);
				\draw (0.5,-2.25) -- +(0.2,-0.5);
				\draw (0.5,-2.25) -- +(-0.2,-0.5);
				\node at (0.5,-3) {o};
				
				\draw[->] (0.07,-0.25) -- (0.07,0.25);
				\draw[<-] (-0.07,-0.25) -- (-0.07,0.25);
				\node at (0,0.6) {\scalebox{0.7}{$\ket{\psi}_\mathrm{s}=\psi_\uparrow\ket{\uparrow}_\mathrm{s} + \psi_\downarrow\ket{\downarrow}_\mathrm{s}$}};
				
				\draw[->] (1.07,-1.8) -- (1.07,-1.3);
				\draw[<-] (0.93,-1.8) -- (0.93,-1.3);
				\node at (0.5,-3.4) {\scalebox{0.7}{$\ket{\phi}_\mathrm{o}=\phi_\uparrow\ket{\uparrow}_\mathrm{o} + \phi_\downarrow\ket{\downarrow}_\mathrm{o}$}};
				
				\draw (1.5,-0.45) rectangle +(1.5,-0.6);
				\draw[->,thick] (2.25,-1) -- (2.25,-0.5);
				\node at (2.25,-1.25) {e};
				
				\node at (2.25,-0.15) {\scalebox{0.7}{$\ket{\chi}_\mathrm{e} = \ket{\uparrow}_\mathrm{e}$}};
		\end{tikzpicture}}
		\caption{\label{fig:signal-observer-environmentup} A quantum system (the signal) is measured by an observer. Unitarity of quantum mechanics necessitates the involvement of another system (the environment) in order to facilitate quantum measurement.}
	\end{figure}

	In order for the measurement procedure to be unitary it is required that the information in the observer (equation \eqref{eq:observer-initial}) be transferred to another system so that the no-deletion theorem \cite{no-deletion} is not violated. It is therefore required to consider a third system into which the observer's prior state can be dumped\footnote{The environment Hilbert space has to be large enough to accomodate the observer's state. Incidentally, the observer state itself might live in a larger Hilbert space than the signal, in other words, there may correspond physical configurations of the observer that do not correspond to having measured any state of the signal. This is not an issue as any arbitrary initial observer state can be dumped into an environment which is large enough.}. We call this the environment and consider for its state
	\begin{equation}
		\ket{\chi}_\mathrm{e}=\ket{\uparrow}_\mathrm{e}.
	\end{equation}
	In this case the state is not in a superposition because we assume an environment state that is well known. The extension to an arbitrary environment state is considered later in subsection \ref{sec:unitarity:generic}. Figure \ref{fig:signal-observer-environmentup} summarises the setup we consider here.
	
	\subsection{\label{sec:unitarity:measurement} Quantum Measurement}
	We define quantum measurement as a process that establishes perfect correlation between the signal and the observer. We restrict ourselves to the simplified and idealised case where the outcomes of the quantum measurement are orthogonal states for both the system and observer; the case of generalised measurements is considered in \cite{measurement-unitary}. In this case, perfect correlation means that the different states of the observer correspond bijectively to states of the signal; cases where the "basis" for the observer differs slightly from that of the signal is considered in \cite{measurement-unitary}. Therefore, starting with the three disconnected subsystem of figure \ref{fig:signal-observer-environmentup} the goal is to reach a state where the signal and observer completely agree with one another
	\begin{eqnarray}
		\ket{\psi}_\mathrm{s}\ket{\phi}_\mathrm{o}\ket{\chi}_\mathrm{e} &=& (\psi_\uparrow\ket{\uparrow} + \psi_\downarrow\ket{\downarrow})_\mathrm{s}(\phi_\uparrow\ket{\uparrow} + \phi_\downarrow\ket{\downarrow})_\mathrm{o}\ket{\uparrow}_\mathrm{e} \nonumber \\
		&\to& (\psi_\uparrow\ket{\uparrow\uparrow} + \psi_\downarrow\ket{\downarrow\downarrow})_{\mathrm{so}}(\phi_\uparrow\ket{\uparrow}+\phi_\downarrow\ket{\downarrow})_\mathrm{e} \\
		&=:& \ket{\Psi}_\mathrm{so}\ket{\phi}_\mathrm{e}; \nonumber
	\end{eqnarray}
	the perfectly correlated signal-observer state is denoted as $\ket{\Psi}_\mathrm{so} = \psi_\uparrow\ket{\uparrow\uparrow}_\mathrm{so} + \psi_\downarrow\ket{\downarrow\downarrow}_\mathrm{so}$.
	
	At the level of the individual basis vectors the action is
	\begin{eqnarray} \label{eq:signal-observer-correlated}
		\ket{\uparrow\uparrow{\uparrow}}_\mathrm{soe} &\to& \ket{{\uparrow\uparrow}\uparrow}_\mathrm{soe} \nonumber \\
		\ket{\uparrow\downarrow{\uparrow}}_\mathrm{soe} &\to& \ket{{\uparrow\uparrow}\downarrow}_\mathrm{soe} \nonumber \\
		\ket{\downarrow\uparrow{\uparrow}}_\mathrm{soe} &\to& \ket{{\downarrow\downarrow}\uparrow}_\mathrm{soe} \\
		\ket{\downarrow\downarrow{\uparrow}}_\mathrm{soe} &\to& \ket{{\downarrow\downarrow}\downarrow}_\mathrm{soe}. \nonumber
	\end{eqnarray}
	
	\subsection{\label{sec:unitarity:generic} Generic Environment}
	In case the environment is not in the state $\ket{\uparrow}_\mathrm{e}$ but $\ket{\color{MPIred} \downarrow}_\mathrm{e}$, by the requirement of unitarity, the subspace $\ket{XY{\color{MPIred} \downarrow}}_\mathrm{soe}$ can only be mapped to span$\{\ket{\uparrow\downarrow Z}_\mathrm{soe}, \ket{\downarrow\uparrow Z}_\mathrm{soe}\}$ so that the measurement procedure necessarily goes wrong; that is, the signal and observer are no longer correlated but anticorrelated.
	
	In terms of the basis vectors a natural extension of the action of equation \eqref{eq:signal-observer-correlated} is
	\begin{eqnarray}
		\ket{\uparrow\uparrow{\color{MPIred}\downarrow}}_\mathrm{soe} &\to& \ket{{\uparrow\downarrow}\uparrow}_\mathrm{soe} \nonumber \\
		\ket{\uparrow\downarrow{\color{MPIred}\downarrow}}_\mathrm{soe} &\to& \ket{{\uparrow\downarrow}\downarrow}_\mathrm{soe} \nonumber \\
		\ket{\downarrow\uparrow{\color{MPIred}\downarrow}}_\mathrm{soe} &\to& \ket{{\downarrow\uparrow}\uparrow}_\mathrm{soe} \\
		\ket{\downarrow\downarrow{\color{MPIred}\downarrow}}_\mathrm{soe} &\to& \ket{{\downarrow\uparrow}\downarrow}_\mathrm{soe} \nonumber
	\end{eqnarray}
	where the environment absorbs the state the observer initially is in. Therefore, the measurement procedure proceeds as
	\begin{equation}
		\ket{\psi}_\mathrm{s}\ket{\phi}_\mathrm{o}\ket{{\color{MPIred}\downarrow}}_\mathrm{e} \to (\psi_\uparrow\ket{\uparrow\downarrow} + \psi_\downarrow\ket{\downarrow\uparrow})_\mathrm{so}\ket{\phi}_\mathrm{e} =: \ket{\bar{\Psi}}_\mathrm{so}\ket{\phi}_\mathrm{e}
	\end{equation}
	where $\ket{\bar{\Psi}}_\mathrm{so} = (\psi_\uparrow\ket{\uparrow\downarrow} + \psi_\downarrow\ket{\downarrow\uparrow})_\mathrm{so}$ is the state of signal-observer where they are anticorrelated.
	
	In case the environment state is in a generic superposition such as $\ket{\chi}_\mathrm{e} = \chi_\uparrow\ket{\uparrow} + {\color{MPIred}\chi_\downarrow}\ket{{\color{MPIred}\downarrow}}$ the measurement proceeds as
	\begin{equation} \label{eq:overall-measurement}
		\ket{\psi}_\mathrm{s}\ket{\phi}_\mathrm{o}(\chi_\uparrow\ket{\uparrow} + {\color{MPIred}\chi_\downarrow}\ket{{\color{MPIred}\downarrow}})_\mathrm{e} \to (\chi_\uparrow\ket{\Psi}_\mathrm{so} + {\color{MPIred}\chi_\downarrow}\ket{\bar{\Psi}}_\mathrm{so})\ket{\phi}_\mathrm{e},
	\end{equation}
	where the signal-observer is correlated in the branch beginning with $\ket{\uparrow}_\mathrm{e}$ and anticorrelated in the other branch, $\ket{\color{MPIred} \downarrow}_\mathrm{e}$.
	
	\subsection{\label{sec:unitarity:local} Local Operations}
	The results of section \ref{sec:unitarity:generic} can be obtained using only local operators. The local operators codify interactions that take place between systems which are physically close to each other. Using local operators between observer-environment and signal-environment, the measurement procedure of equation \eqref{eq:overall-measurement} can be carried out; there is no requirement for the signal and observer to directly interact.
	
	An imprint operation is defined as
	\begin{align} \label{eq:imprint-operation}
		\mathcal{I}_{\mathrm{a} \to \mathrm{b}}\ket{\uparrow\uparrow}_\mathrm{ab} &= \ket{\uparrow\uparrow}_\mathrm{ab} \nonumber \\
		\mathcal{I}_{\mathrm{a} \to \mathrm{b}}\ket{\uparrow\downarrow}_\mathrm{ab} &= \ket{\uparrow\downarrow}_\mathrm{ab} \nonumber \\
		\mathcal{I}_{\mathrm{a} \to \mathrm{b}}\ket{\downarrow\uparrow}_\mathrm{ab} &= \ket{\downarrow\downarrow}_\mathrm{ab} \\
		\mathcal{I}_{\mathrm{a} \to \mathrm{b}}\ket{\downarrow\downarrow}_\mathrm{ab} &= \ket{\downarrow\uparrow}_\mathrm{ab} \nonumber
	\end{align}
	and is similar to the CNOT gate used in quantum computation. A swap operation is defined as
	\begin{align} \label{eq:swap-operation}
		\mathcal{S}_{\mathrm{a} \leftrightarrow \mathrm{b}}\ket{\uparrow\uparrow}_\mathrm{ab} &= \ket{\uparrow\uparrow}_\mathrm{ab} \nonumber \\
		\mathcal{S}_{\mathrm{a} \leftrightarrow \mathrm{b}}\ket{\uparrow\downarrow}_\mathrm{ab} &= \ket{\downarrow\uparrow}_\mathrm{ab} \nonumber \\
		\mathcal{S}_{\mathrm{a} \leftrightarrow \mathrm{b}}\ket{\downarrow\uparrow}_\mathrm{ab} &= \ket{\uparrow\downarrow}_\mathrm{ab} \\
		\mathcal{S}_{\mathrm{a} \leftrightarrow \mathrm{b}}\ket{\downarrow\downarrow}_\mathrm{ab} &= \ket{\downarrow\downarrow}_\mathrm{ab} \nonumber
	\end{align}
	which is similar to the swap gate in quantum computing. It can now readily be seen that $\mathcal{S}_{o \leftrightarrow e} \circ \mathcal{I}_{s \rightarrow e}$ achieves the required outcome of equation \eqref{eq:overall-measurement},
	\begin{equation}
		\mathcal{S}_{o \leftrightarrow e} \circ \mathcal{I}_{s \rightarrow e} \quad \left[\ket{\psi}_\mathrm{s}\ket{\phi}_\mathrm{o}(\chi_\uparrow\ket{\uparrow} + {\color{MPIred}\chi_\downarrow}\ket{{\color{MPIred}\downarrow}})_\mathrm{e}\right] \to (\chi_\uparrow\ket{\Psi}_\mathrm{so} + {\color{MPIred}\chi_\downarrow}\ket{\bar{\Psi}}_\mathrm{so})\ket{\phi}_\mathrm{e}.
	\end{equation}
	
\begin{figure}[h]
		\centering
		\resizebox{6cm}{!}{\begin{tikzpicture}
				\node at (-1,2.4) {\Large s:};
				\node at (-1,1.2) {\Large o:};
				\node at (-1,0) {\Large e:};
				
				\node[cloud, aspect=1.2, fill=MPIgreen!50] at (0,2.4) {$\ket{i}$};
				\node[cloud, aspect=1.2, fill=MPIdarkyellow] at (0,1.2) {$\ket{\phi}$};
				\node[cloud, aspect=1.2, fill=MPIred!50] at (0,0) {$\ket{k}$};
				
				\node at (3,1.2) {$\longrightarrow$};
				
				\node[cloud, aspect=1.2, fill=MPIgreen!50] at (6,2.4) {$\ket{i}$};
				\node[cloud, aspect=1.6, fill=MPIred!50!MPIgreen!50] at (6,1.5) {$\ket{k \circ i}$};
				\node[cloud, aspect=1.2, fill=MPIdarkyellow] at (6,0) {$\ket{\phi}$};
		\end{tikzpicture}}
		\caption{\label{fig:uncorrected-measurement-procedure} Measurement procedure without environmental correction. The environment influences the measurement procedure as indicated by the term $k \circ i$. The clouds indicate systems which are correlated with other. Two clouds touching indicates that the quantum systems they are part of are correlated (and thereby entangled). The clouds are coloured to distinguish between the different quantum systems participating in the measurement procedure.}
	\end{figure}
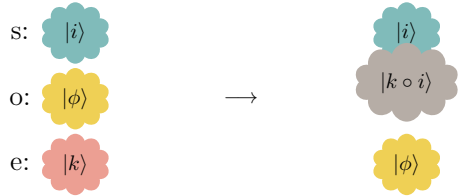

	The environmental influence on the measurement procedure is evidenced by the fact that there are multiple branches of the wavefunction (equation \eqref{eq:overall-measurement}) depending on which branch of the environment the overall system began in. The signal-observer are perfectly correlated in one of the branches and anticorrelated in the other. In the following, section \ref{sec:resource}, we explore how this environmental influence can be corrected for. The measurement procedure so far is summarised in figure \ref{fig:uncorrected-measurement-procedure}.
	
	\section{\label{sec:resource} Correlation as a Resource}
	In this section we discuss how the influence of the environment on the measurement procedure can be corrected for. There must be a redundant record of the environment in order for this correction to work. We argue that a correlated environment of the form
	\begin{equation} \label{eq:correlated-environment}
		\ket{\chi}_\mathrm{e} = \sum_k \chi_k \ket{k}_\mathrm{e_1}\ket{k}_\mathrm{e_2}\ket{k}_\mathrm{e_3}...\ket{k}_\mathrm{e_N}
	\end{equation}
	is used to correct for the environmental influence.
	
	\subsection{\label{sec:resource:procedure} Measurement Procedure}
	The measurement procedure proceeds as follows, the observer first swaps its state with the environment using a swap operation. This makes the observer now part of the correlated environment. Therefore, the environmental state can be corrected to obtain a "0 state" $\ket{\uparrow}_\mathrm{o}$ thereby allowing the usual imprint operation by signal onto the observer to result in a perfectly correlated state. Note that this decoherence of the observer into the network of the environment could take place long before the signal is measured. It enables the observer to measure the signal and also sets the basis of measurement. The resulting state is as in figure \ref{fig:corrected-measurement-procedure}.
	
	The mathematics works as follows. The swap operation enables the observer to become a part of the correlated environment
	\begin{align} \label{eq:corrected-measurement-procedure-1}
		&\mathrm{s}: \ket{i}_\mathrm{s} & & & &\mathrm{s}: \ket{i}_\mathrm{s} \nonumber \\
		{\color{MPIred}\mathcal{S}_{\mathrm{o} \to \mathrm{e_N}}}\quad &\mathrm{o}: {\color{MPIred}\ket{\phi}_\mathrm{o}} & &\to & &\mathrm{o}: \ket{k}_\mathrm{o} \\
		&\mathrm{e}: \ket{k}_\mathrm{e_1}\ket{k}_\mathrm{e_2}\dots{\color{MPIred}\ket{k}_\mathrm{e_N}} & & & &\mathrm{e}: \ket{k}_\mathrm{e_1}\ket{k}_\mathrm{e_2}\dots\ket{\phi}_\mathrm{e_N}. \nonumber
	\end{align}
	The environmental influence can now be corrected for
	\begin{align} \label{eq:corrected-measurement-procedure-2}
		&\mathrm{s}: \ket{i}_\mathrm{s} & & & &\mathrm{s}: {\color{MPIred}\ket{i}_\mathrm{s}} \nonumber \\
		{\color{MPIred}\mathcal{I}^{-1}_\mathrm{e_2 \to e_1}} \quad &\mathrm{o}: \ket{k}_\mathrm{o} & &\to & {\color{MPIred}\mathcal{I}_\mathrm{s \to e_1}} \quad &\mathrm{o}: \ket{k}_\mathrm{o} \nonumber \\
		&\mathrm{e}: {\color{MPIred}\ket{k}_\mathrm{e_1}}{\color{MPIred}\ket{k}_\mathrm{e_2}}\dots\ket{\phi}_\mathrm{e_N} & & & &\mathrm{e}: {\color{MPIred}\ket{\uparrow}_\mathrm{e_1}}\ket{k}_\mathrm{e_2}\dots\ket{\phi}_\mathrm{e_N} \nonumber \\
		& & & & & \nonumber \\
		& & & & &\downarrow \\
		& & & & & \nonumber \\
		&\mathrm{s}: \ket{i}_\mathrm{s} & & & &\mathrm{s}: \ket{i}_\mathrm{s} \nonumber \\
		&\mathrm{o}: \ket{i}_\mathrm{o} & &\leftarrow & {\color{MPIred}\mathcal{S}_\mathrm{s \leftrightarrow e_1}} \quad & \mathrm{o}: {\color{MPIred}\ket{k}_\mathrm{o}} \nonumber \\
		&\mathrm{e}: \ket{k}_\mathrm{e_1}\ket{k}_\mathrm{e_2}\dots\ket{\phi}_\mathrm{e_N} & & & &\mathrm{e}: {\color{MPIred}\ket{i}_\mathrm{e_1}}\ket{k}_\mathrm{e_2}\dots\ket{\phi}_\mathrm{e_N} \nonumber
	\end{align}
	and perfect correlation between signal and observer results. $\mathcal{I}^{-1}_\mathrm{a \to b}$ is the inverse of the imprint operation; this action undoes an imprint, $\mathcal{I}_\mathrm{a \to b}$, that system a might have made on system b. In case of the CNOT gate of equation \eqref{eq:imprint-operation}, the inverse imprint is the same as imprint.
	
	\subsection{\label{sec:resource:correlation} Correlation as a Resource}
\begin{figure}[h]
		\centering
		\resizebox{11cm}{!}{\begin{tikzpicture}
				\node[cloud, fill=MPIgreen!50, aspect=1.5, cloud puffs = 15] at (0,0) {\phantom{reeeeee}};
				\node[cloud, fill=MPIdarkyellow, aspect=1.5, cloud puffs = 15] at (1.25,-1.25) {\phantom{reeee}};
				\node[cloud, fill=MPIred!50, aspect=3, cloud puffs = 18] at (4.375,0) {\phantom{reeeeeeeeeeeeeeeeeeee}};
				
				\node[block] (sys) at (0,0) {s};
				\node[block] (obs) at (1.25,-1.25) {o};
				\node[block] (env1) at (2.5,0) {$e_1$};
				\node[block] (env2) at (3.75,0) {$e_2$};
				\node[block] (env3) at (5,0) {...};
				\node[block] (env4) at (6.25,0) {$e_N$};
				
				\node at (4.375,1.5) {\Large N - 1};
				
				\node at (8.125,-0.75) {$\to$};
				
				\node[cloud, fill=MPIgreen!50, aspect=1.5, cloud puffs = 15] at (10,0) {\phantom{reeeeee}};
				\node[cloud, fill=MPIgreen!50, aspect=1.5, cloud puffs = 15, rotate = -15] at (11,-1.25) {\phantom{reeeeee}};
				\node[cloud, fill=MPIdarkyellow, aspect=1.5, cloud puffs = 15] at (17,0) {\phantom{reeee}};
				\node[cloud, fill=MPIred!50, aspect=2.3, cloud puffs = 18] at (13.75,0) {\phantom{reeeeeeeeeeeeee}};
				
				\node[block] (sys') at (10,0) {s};
				\node[block] (obs') at (11.25,-1.25) {o};
				\node[block] (env1') at (12.5,0) {$e_1$};
				\node[block] (env2') at (13.75,0) {$e_2$};
				\node[block] (env3') at (15,0) {...};
				\node[block] (env4') at (17,0) {$e_N$};
				
				\node at (10,1.5) {\Large 1};
				\node at (13.75,1.5) {\Large N - 2};
		\end{tikzpicture}}
		\caption{\label{fig:corrected-measurement-procedure} Measurement procedure with environmental correction. The redundant information provided by the correlated environment is used for this correction. Also indicated are the measures of correlation in the involved subsystems.}
	\end{figure}
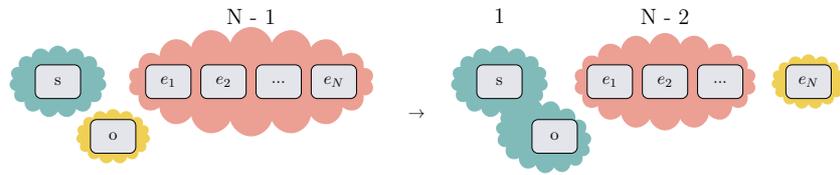

	We propose that there is a cost associated with correlating a signal-observer pair as the environment loses some of its own correlation; it is a resource used up in the measurement procedure. By associating a well motivated measure it can be seen that the resource of correlation is conserved over the measurement procedure, as described in figure \ref{fig:corrected-measurement-procedure}.
	
	For every signal-observer qubit pair that a correlated environment in turn correlates, it must lose at least one qubit of correlation. This is because of the basic requirement of the observer having to dump its arbitrary initial state somewhere due to constraints of the no-deletion theorem. Therefore, assuming a completely efficient procedure, the cost of correlating one signal-observer pair is one qubit of correlation. 
	
	Proceeding this way it can be seen that a correlated environment as in equation \eqref{eq:correlated-environment} can correlate a further N-1 signal-observer pairs before it runs out of all its ability to correlate. Therefore, for a perfectly efficient procedure the measure of correlation of a correlated environment state as in equation \eqref{eq:correlated-environment} is N-1.
	
	A newly correlated signal-observer pair can itself sacrifice its correlation to correlate a different signal-observer pair. Therefore, the resource of correlation does not vanish, but is conserved and transferred from the environment to the newly correlated signal-observer pair.
	
	\section{\label{sec:objective} Objective Classical Reality}
	In this section we explore how the objective classical reality we are all familiar with emerges from the considerations in the previous sections, \ref{sec:unitarity} and \ref{sec:resource}.
	
	\subsection{\label{sec:objective:multiple} Multiple Observers}
	If multiple observers perform the above measurement procedure as in figure \ref{fig:corrected-measurement-procedure}, the result is a network of observers 
	\begin{eqnarray}
		& & \left(\sum_{i} \psi_i \ket{i}_\mathrm{s}\right)\ket{\phi_1}_\mathrm{o_1}\ket{\phi_2}_\mathrm{o_2}\ket{\phi_3}_\mathrm{o_3}... \nonumber \\
		&\overset{\text{o$_1$ measures}}{\to}& \left(\sum_{i} \psi_i \ket{i}_\mathrm{s}\ket{i}_\mathrm{o_1}\right)\ket{\phi_2}_\mathrm{o_2}\ket{\phi_3}_\mathrm{o_3}... (\text{environment suppressed}) \nonumber \\
		&\overset{\text{o$_2$ measures}}{\to}& \left(\sum_{i} \psi_i \ket{i}_\mathrm{s}\ket{i}_\mathrm{o_1}\ket{i}_\mathrm{o_2}\right)\ket{\phi_3}_\mathrm{o_3}... \\
		&\overset{...}{\to}& \left(\sum_{i} \psi_i \ket{i}_\mathrm{s}\ket{i}_\mathrm{o_1}\ket{i}_\mathrm{o_2}\ket{i}_\mathrm{o_3}...\right). \nonumber 
	\end{eqnarray}
	This enables the repeatability of experiment and, therefore, for multiple observers to agree with one another on what the state of reality of a signal is. This is the definition of objectivity used in the current article, multiple entities all in agreement with one another about the state of a signal (in agreement with \cite{spectrum-broadcast}).
	
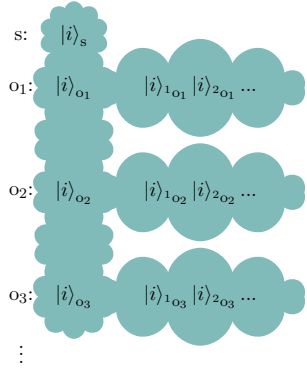
\begin{figure}[h]
		\centering
		\resizebox{4cm}{!}{\begin{tikzpicture}
				\node at (0,0) {s:};
				\node at (0,-1) {o$_1$:};
				\node at (0,-3) {o$_2$:};
				\node at (0,-5) {o$_3$:};
				
				\node[cloud, aspect=1, fill=MPIgreen!50] at (1,0) {$\ket{i}_\mathrm{s}$};
				\node[cloud, aspect=1, fill=MPIgreen!50] at (1,-1) {$\ket{i}_\mathrm{o_1}$};
				\node[cloud, aspect=4, fill=MPIgreen!50] at (3.4,-1) {$\ket{i}_\mathrm{^1o_1}\ket{i}_\mathrm{^2o_1}...$};
				\node[cloud, aspect=1, fill=MPIgreen!50] at (1,-2) {$\phantom{\ket{i}_\mathrm{s_1}}$};
				\node[cloud, aspect=1, fill=MPIgreen!50] at (1,-3) {$\ket{i}_\mathrm{o_2}$};
				\node[cloud, aspect=4, fill=MPIgreen!50] at (3.4,-3) {$\ket{i}_\mathrm{^1o_2}\ket{i}_\mathrm{^2o_2}...$};
				\node[cloud, aspect=1, fill=MPIgreen!50] at (1,-4) {$\phantom{\ket{i}_\mathrm{s_1}}$};
				\node[cloud, aspect=1, fill=MPIgreen!50] at (1,-5) {$\ket{i}_\mathrm{o_3}$};
				\node[cloud, aspect=4, fill=MPIgreen!50] at (3.4,-5) {$\ket{i}_\mathrm{^1o_3}\ket{i}_\mathrm{^2o_3}...$};
				\node at (0,-6) {$\vdots$};
		\end{tikzpicture}}
		\caption{\label{fig:classical-reality} A highly branched network of decohered states lends stability to the measurement of the signal as there is a large amount of redundancy in the information. Also, for this information to be deleted all the involved systems must conspire to come together and undo this correlation.}
	\end{figure}

	Furthermore, if each of these observers o$_i$ are themselves classical objects that decohere, what obtains is a highly branched network of states that all agree with one another (figure \ref{fig:classical-reality}). This lends stability to the decohered network of states as in order to delete the information about the state of the signal, all the systems it has interacted with have to come together to conspire to undo this correlation. In the following subsection \ref{sec:objective:different} we consider what happens when measurements are performed in different bases. It turns out that an objective classical reality does not then emerge as the different observers do not agree with one another on what the state of a signal is. The network of redundant information in figure \ref{fig:classical-reality}, is exactly what protects against such a loss of objectivity.
	
	\subsection{\label{sec:objective:different} Different Basis}
	In the discussion of the measurement procedure in sections \ref{sec:unitarity} and \ref{sec:resource} most of the analysis was (made) basis independent. None among signal, observer and environment need be in any particular state. The swap operation of equation \eqref{eq:swap-operation} can also be shown to be basis independent. Explicit dependence on a particular basis only makes its way in equation \eqref{eq:imprint-operation} through the imprint operation. In this subsection we explore what happens in case the measurement is performed in a different basis. It shall be seen that the measurement procedure need no longer give rise to an objective classical reality, different observers may disagree with one another on the state of a signal.
	
\begin{figure}[h]
		\centering
		\resizebox{10cm}{!}{\begin{tikzpicture}			
				\node[cloud, aspect=1, fill=MPIgreen!50, opacity=0.5] at (0,0) {\phantom{eeee}};
				\node[block] at (0,0) {s};
				
				\node at (1.25,0) {$\to$};
				
				\node[cloud, aspect=3, fill=MPIgreen!50, opacity=0.5] at (3.25,0) {\phantom{eeeeeeeeeeee}};
				\node[block] at (2.5,0) {s};
				\node[block] at (4,0) {o$_1$};
				
				\node at (5.25,0) {$\to$};
				
				\node[cloud, aspect=3, fill=MPIgreen!50, opacity=0.5] at (7.25,0) {\phantom{eeeeeeeeeeee}};
				\node[cloud, aspect=0.4, fill=MPIred!50, opacity=0.5] at (6.5,-0.5) {\phantom{eee}};
				\node[block] at (6.5,0) {s};
				\node[block] at (8,0) {o$_1$};
				\node[block] at (6.5,-1) {o$_2$};
				
				\node at (9.25,0) {$\to$};
				
				\node[cloud, aspect=3, fill=MPIgreen!50, opacity=0.5] at (11.25,0) {\phantom{eeeeeeeeeeee}};
				\node[cloud, aspect=0.4, fill=MPIred!50, opacity=0.5] at (10.5,-0.5) {\phantom{eee}};
				\node[cloud, aspect=0.4, fill=MPIdarkyellow, opacity=0.5] at (12,-0.5) {\phantom{eee}};
				\node[block] at (10.5,0) {s};
				\node[block] at (12,0) {o$_1$};
				\node[block] at (10.5,-1) {o$_2$};
				\node[block] at (12,-1) {o$'_3$};
		\end{tikzpicture}}
		\caption{\label{fig:different-basis} Measurement in a different basis is no longer objective. Different observers may disagree on what consititutes reality. Transparent clouds over the same quantum system indicate its being observed by observers in different bases.}
	\end{figure}
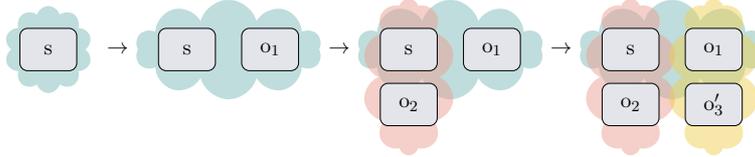

	Consider a signal s that is first measured by an observer o$_1$ in a particular basis $\{\ket{\uparrow}, \ket{\downarrow}\}$. Let it further be observed by a different observer o$_2$ in a different basis
	\begin{align}
		\ket{\rightarrow} &= \frac{1}{\sqrt{2}}(\ket{\uparrow} + \ket{\downarrow}) & \ket{\leftarrow} &= \frac{1}{\sqrt{2}}(\ket{\uparrow} - \ket{\downarrow}).
	\end{align}
	Let the observer o$_1$ further be measured by a third observer o$'_3$ in the new basis. The resulting expressions are (ignoring normalisation of the quantum state):
	\begin{eqnarray} \label{eq:different-basis}
		&& \psi_\uparrow \ket{\uparrow}_\mathrm{s} + \psi_\downarrow \ket{\downarrow}_\mathrm{s} \nonumber \\
		&\to& \psi_\uparrow \ket{\uparrow}_\mathrm{s}\ket{\uparrow}_\mathrm{o_1} + \psi_\downarrow \ket{\downarrow}_\mathrm{s}\ket{\downarrow}_\mathrm{o_1} \nonumber \\
		&\to& \ket{\rightarrow\rightarrow}_\mathrm{so_2}\left(\psi_\uparrow\ket{\uparrow}_\mathrm{o_1} + \psi_\downarrow\ket{\downarrow}_\mathrm{o_1}\right) +  \ket{\leftarrow\leftarrow}_\mathrm{so_2}\left(\psi_\uparrow\ket{\uparrow}_\mathrm{o_1} - \psi_\downarrow\ket{\downarrow}_\mathrm{o_1}\right) \\
		&\to& (\psi_\uparrow + \psi_\downarrow)\ket{\rightarrow\rightarrow}_\mathrm{so_2}\ket{\rightarrow\rightarrow}_\mathrm{o_1o'_3} + (\psi_\uparrow - \psi_\downarrow){\color{MPIred} \ket{\leftarrow\leftarrow}_\mathrm{so_2}\ket{\rightarrow\rightarrow}_\mathrm{o_1o'_3}} \nonumber \\
		&& + (\psi_\uparrow - \psi_\downarrow){\color{MPIred} \ket{\rightarrow\rightarrow}_\mathrm{so_2}\ket{\leftarrow\leftarrow}_\mathrm{o_1o'_3}} + (\psi_\uparrow + \psi_\downarrow)\ket{\leftarrow\leftarrow}_\mathrm{so_2}\ket{\leftarrow\leftarrow}_\mathrm{o_1o'_3} \nonumber
	\end{eqnarray}
	and it is seen that there are instances (those highlighted in {\color{MPIred} red} in equation \eqref{eq:different-basis}) where the different observers o$_2$ and o$'_3$ do not agree on the state of the signal. Objectivity is lost! Figure \ref{fig:different-basis} summarises the argument.
	
	However, in case the observer o$_1$ was classical it is further observed by other observers and this creates a network of observers which allows recovery of the information regarding measurement of the signal (figure \ref{fig:different-basis-decohered}). The mathematics is worked out in appendix \ref{app:different-basis-decohered}.
	\begin{figure}[h]
		\centering
		\resizebox{10cm}{!}{\begin{tikzpicture}
				\node[cloud, aspect=1, fill=MPIgreen!50, opacity=0.5] at (0,0) {\phantom{eeee}};
				\node[block] at (0,0) {s};
				
				\node at (1.25,0) {$\to$};
				
				\node[cloud, aspect=3, fill=MPIgreen!50, opacity=0.5] at (3.25,0) {\phantom{eeeeeeeeeeee}};
				\node[cloud, aspect=0.3, fill=MPIgreen!50, opacity=0.5] at (4,1.5) {\phantom{eee}};
				\node[block] at (2.5,0) {s};
				\node[block] at (4,0) {o$_1$};
				\node[block] at (4,1) {$^1$o$_1$};
				\node[block] at (4,2) {$\vdots$};
				
				\node at (5.25,0) {$\to$};
				
				\node[cloud, aspect=3, fill=MPIgreen!50, opacity=0.5] at (7.25,0) {\phantom{eeeeeeeeeeee}};
				\node[cloud, aspect=0.4, fill=MPIred!50, opacity=0.5] at (6.5,-0.5) {\phantom{eee}};
				\node[cloud, aspect=0.3, fill=MPIgreen!50, opacity=0.5] at (8,1.5) {\phantom{eee}};
				\node[block] at (6.5,0) {s};
				\node[block] at (8,0) {o$_1$};
				\node[block] at (8,1) {$^1$o$_1$};
				\node[block] at (8,2) {$\vdots$};
				\node[block] at (6.5,-1) {o$_2$};
				
				\node at (9.25,0) {$\to$};
				
				\node[cloud, aspect=3, fill=MPIgreen!50, opacity=0.5] at (11.25,0) {\phantom{eeeeeeeeeeee}};
				\node[cloud, aspect=0.4, fill=MPIred!50, opacity=0.5] at (10.5,-0.5) {\phantom{eee}};
				\node[cloud, aspect=0.4, fill=MPIdarkyellow, opacity=0.5] at (12,-0.5) {\phantom{eee}};
				\node[cloud, aspect=0.3, fill=MPIgreen!50, opacity=0.5] at (12,1.5) {\phantom{eee}};
				\node[block] at (10.5,0) {s};
				\node[block] at (12,0) {o$_1$};
				\node[block] at (12,1) {$^1$o$_1$};
				\node[block] at (12,2) {$\vdots$};
				\node[block] at (10.5,-1) {o$_2$};
				\node[block] at (12,-1) {o$'_3$};
		\end{tikzpicture}}
		\caption{\label{fig:different-basis-decohered} Due to the redundant information stored in the environment it is possible to recover information about the environment and thereby allow for the signal to once again be determined objectively.}
	\end{figure}
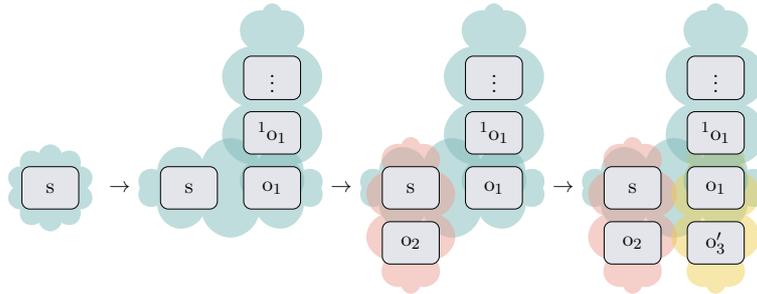
	
	It could be argued that what is made objective is not the state of the signal but rather the measurement of the signal by the observer. In this sense, it is the whole network of observer states that is assigned the adjective "objective". This, of course, implies that the state of the signal is also objective in the sense that there is now a redundant record of the measurement as in figure \ref{fig:classical-reality} such that all observers within that network agree with one another on its state. Moreover, this information is safeguarded against measurements of the signal in different bases due to the redundancy of information stored in the network of observers. In this sense, the network of figure \ref{fig:different-basis} is unstable whereas that of figure \ref{fig:different-basis-decohered} is stable.
	
	\section{\label{sec:discussion} Discussion}
	We begin with an assumption of purely unitary quantum dynamics in section \ref{sec:unitarity} and explained the limits it imposes on the quantum measurement procedure. The assumption of unitarity along with that of arbitrary initial state for signal and observer necessitates the use of an environment into which the observer can dispose its initial state. Extending the arbitrariness of initial state to the environment (and along with the usual assumption of unitarity) it is seen that the environment can influence the measurement procedure and leave a trace therein (equation \eqref{eq:overall-measurement}).
	
	This implies the necessity of redundant information about the environment so that the influence of the environment can be undone. We therefore describe a correlated environment (equation \eqref{eq:correlated-environment}) which is redundantly encoded and which can help correct for the environmental influence. With this addition we describe a corrected measurement procedure (equations \eqref{eq:corrected-measurement-procedure-1} and \eqref{eq:corrected-measurement-procedure-2}) which results in perfect correlation between signal and observer. It is also seen that one can come up with a measure of this correlation and that it is a conserved resource used up in the quantum measurement procedure. These considerations make up section \ref{sec:resource}.
	
	Finally, in section \ref{sec:objective} we describe how, using the considerations of the previous sections, the emergence of an objective classical reality can be argued. It is seen that redundancy of information about the measurement procedure is key, as it is the agreement of different observers regarding the state of affairs of the world that is described as objective classical reality \cite{measurement-unitary, zurek-classical, spectrum-broadcast}. Stability of this classical reality is argued to be due to the formation of highly branched structures such as in figure \ref{fig:classical-reality} which makes it difficult to undo this correlation.
	
	The current article discusses a particular example in detail in order to highlight the main ideas of our approach. A related article \cite{measurement-unitary} is a more abstract analysis and addresses other issues in more detail. One of the issues we do not address in either article is how a correlated environment such as in equation \eqref{eq:correlated-environment} emerges in the first place. While one might obtain a hint of an answer in section \ref{sec:objective}, detailed analysis is required in order to properly address this issue. The issue of basis independence is also something we gloss over. While the state of signal, observer, and environment are all basis independent and so is the swap operation (equation \eqref{eq:swap-operation}), the imprint operation (equation \eqref{eq:imprint-operation}) is basis dependent. Therefore, the issue of preferred basis brought up in section \ref{sec:unitarity} is not completely addressed. How/whether the emergence of basis dependence in the imprint operation relates to the emergence of a correlated environment is another interesting question for future research.
	
	\vspace{6pt}
	
	\appendix
	\section[Appendix ~\thesection]{Appendix} \label{app:different-basis-decohered}
	The mathematics following figure \ref{fig:different-basis-decohered} proceeds as follows
	\begin{equation} \label{eq:different-basis-decohered}
		\psi_\uparrow \ket{\uparrow}_\mathrm{s} + \psi_\downarrow \ket{\downarrow}_\mathrm{s} \to \psi_\uparrow \ket{\uparrow}_\mathrm{s}\overset{\overset{\vdots}{\ket{\uparrow}_\mathrm{^1o_1}}}{\ket{\uparrow}_\mathrm{o_1}} + \psi_\downarrow \ket{\downarrow}_\mathrm{s}\overset{\overset{\vdots}{\ket{\downarrow}_\mathrm{^1o_1}}}{\ket{\downarrow}_\mathrm{o_1}}
	\end{equation}
	where the state $\overset{\overset{\vdots}{\ket{i}_\mathrm{^1o_1}}}{\ket{i}_\mathrm{o_1}}$ indicates that the observer o$_1$ is a highly branched state which is observed by several consequent observers. The observation in different bases then proceeds as follows
	\begin{eqnarray}
		&\to& \ket{\rightarrow\rightarrow}_\mathrm{so_2}(\psi_\uparrow\overset{\overset{\vdots}{\ket{\uparrow}_\mathrm{^1o_1}}}{\ket{\uparrow}_\mathrm{o_1}} + \psi_\downarrow\overset{\overset{\vdots}{\ket{\downarrow}_\mathrm{^1o_1}}}{\ket{\downarrow}_\mathrm{o_1}}) +  \ket{\leftarrow\leftarrow}_\mathrm{so_2}(\psi_\uparrow\overset{\overset{\vdots}{\ket{\uparrow}_\mathrm{^1o_1}}}{\ket{\uparrow}_\mathrm{o_1}} - \psi_\downarrow\overset{\overset{\vdots}{\ket{\downarrow}_\mathrm{^1o_1}}}{\ket{\downarrow}_\mathrm{o_1}}) \\
		&\to& (\overset{\overset{\vdots}{\ket{\uparrow}_\mathrm{^1o_1}}}{\psi_\uparrow} + \overset{\overset{\vdots}{\ket{\downarrow}_\mathrm{^1o_1}}}{\psi_\downarrow})\ket{\rightarrow\rightarrow}_\mathrm{so_2}\ket{\rightarrow\rightarrow}_\mathrm{o_1o'_3} +  (\overset{\overset{\vdots}{\ket{\uparrow}_\mathrm{^1o_1}}}{\psi_\uparrow} - \overset{\overset{\vdots}{\ket{\downarrow}_\mathrm{^1o_1}}}{\psi_\downarrow}){\color{MPIred}\ket{\leftarrow\leftarrow}_\mathrm{so_2}\ket{\rightarrow\rightarrow}_\mathrm{o_1o'_3}} \nonumber \\
		&& +  (\overset{\overset{\vdots}{\ket{\uparrow}_\mathrm{^1o_1}}}{\psi_\uparrow} - \overset{\overset{\vdots}{\ket{\downarrow}_\mathrm{^1o_1}}}{\psi_\downarrow}){\color{MPIred}\ket{\rightarrow\rightarrow}_\mathrm{so_2}\ket{\leftarrow\leftarrow}_\mathrm{o_1o'_3}} + (\overset{\overset{\vdots}{\ket{\uparrow}_\mathrm{^1o_1}}}{\psi_\uparrow} + \overset{\overset{\vdots}{\ket{\downarrow}_\mathrm{^1o_1}}}{\psi_\downarrow})\ket{\leftarrow\leftarrow}_\mathrm{so_2}\ket{\leftarrow\leftarrow}_\mathrm{o_1o'_3}.
	\end{eqnarray}
	
	In this case, observers o$_2$ and o$'_3$ continue to disagree about the measured state of the signal. However, the redundant record maintained in the states $\ket{i}_\mathrm{^1o_1}\dots$ allow other independent observers to query and infer the previously measured state of the signal. Therefore, the redundancy in recording the measurement outcomes for observer o$_1$ allows the formation of a stable objective classical reality for the measurement of signal s despite the disagreement between observers o$_2$ and o$'_3$. The network of states $\ket{i}_\mathrm{^1o_1}\dots$ contains redundant information about the measurement of signal s by observer o$_1$. Redundancy is key!
	
	\section*{Acknowledgements}
	We thank our co-author in \cite{measurement-unitary}, Reimar Leike, for his discussion and advice on the background work leading up to this article.
	
	Philipp Frank acknowledges funding through the German Federal Ministry of Education and Research for the project ErUM-IFT: Informationsfeldtheorie für Experimente an Großforschungsanlagen (Förderkennzeichen: 05D23EO1).
	
	\bibliographystyle{plain}
	\bibliography{references}
		
\end{document}